\documentclass[twocolumn,aps,prl,showpacs,floatfix]{revtex4}
\usepackage{graphicx}
\draft
\newcommand{\norm}{\frac{2}{(2\pi)^3}}
\newcommand{\qd}[1]{\left[ #1 \right]}
\newcommand{\td}[1]{\left( #1 \right)}

\newcommand{\itg}[1]{\norm \int d^3k f_{#1}(k)}
\newcommand{\itau}{\itg{\tau}g(k)}
\newcommand{\rz}{\rho_{_0}}
\newcommand{\umd}[1]{ \td{ \frac{1}{2}+x_{#1} } }

\newcommand{\inew}[1]{\mathcal{I}_#1 }

\begin{document}
\preprint{\it Phys. Rev. C (2004) in press.}
\title{Constraining the neutron-proton effective mass splitting in neutron-rich matter}
\author{\bf Bao-An Li\footnote{email: Bali@astate.edu}}
\address{Department of Physics, Arkansas State University\\
State University, Arkansas 72467-0419, USA}

\date{\today}
\begin{abstract}
Within Bombaci's phenomenological single-nucleon potential model we study the
neutron-proton effective mass splitting $m_n^*- m_p^*$ in neutron-rich matter.
It is shown that an effective mass splitting of $m_n^*<m_p^*$ leads to a symmetry potential 
that is inconsistent with the energy dependence of the Lane potential constrained 
by the nucleon-nucleus scattering experimental data.
\end{abstract}
\pacs{25.70.-z, 25.70.Pq., 24.10.Lx}
\maketitle

One of the most fundamental properties characterizing the propagation of a nucleon in nuclear medium is
its effective mass\cite{dob,mahaux,neg81}. The latter describes to leading order 
the effects related to the non-locality of the underlying nuclear effective interactions and 
the Pauli exchange effects in many-fermion systems. In neutron-rich matter, 
an interesting new question arises as to whether the effective mass $m_n^*$ for neutrons 
is higher or lower than that for protons $m_p^*$. Microscopic many-body theories, e.g., 
the Landau-Fermi liquid theory\cite{sjo76} and the Brueckner-Hartree-Fock (BHF) approach\cite{bhf}, 
predict that $m_n^*>m_p^*$ in neutron-rich matter. On the other hand, some effective interactions within 
phenomenological approaches predict the opposite\cite{riz}. Unfortunately, up to now almost 
nothing is known experimentally about the neutron-proton effective mass splitting $m_n^*-m_p^*$ 
in neutron-rich medium. Knowledge about nucleon effective mass in neutron-rich matter is critically 
important for understanding several properties of neutron 
stars\cite{bethe,cop85,far01}. It is also important for the reaction dynamics of nuclear collisions 
induced by radioactive nuclei, such as, the degree and speed of isospin diffusion, the neutron-proton 
differential collective flow as well as the isospin equilibrium\cite{pan92,gale02,riz,lidas}. 
Moreover, it influences the magnitude of shell effects and basic properties of nuclei fare from 
stability\cite{dob}. However, even the sign of the neutron-proton effective mass splitting 
remains rather controversial theoretically. It is thus imperative to constrain the neutron-proton 
effective mass splitting in neutron-rich matter using experimental data. In a recent study by 
Rizzo et al.\cite{riz}, nuclear reactions with radioactive beams were proposed as a means to 
disentangle the sign of the neutron-proton effective mass splitting in neutron-rich matter. 
In this note we take a completely different approach for the same purpose. 
We show that an effective mass splitting of $m_n^*<m_p^*$ leads to a symmetry potential 
that is inconsistent with the energy dependence of the Lane potential constrained by existing
nucleon-nucleus scattering data. The effective mass splitting of $m_n^*<m_p^*$ is thus ruled out 
phenomenologically using experimental data indirectly.

The effective mass $m_{\tau}^*$ of a nucleon $\tau$ (n or p) is determined by the 
momentum-dependent single nucleon potential $U_{\tau}$ via\cite{mahaux}
\begin{equation}
\frac{m^*_\tau}{m_{\tau}}=\left\{1+
\frac{m_{\tau}}{\hbar^2k}\frac{dU_\tau}{dk}
\right\}
^{-1}_{k=k_{\tau}^F},
\label{mstar}
\end{equation}
where $k_{\tau}^F$ is the nucleon Fermi wave number 
in asymmetric nuclear matter of isospin asymmetry 
$\delta\equiv (\rho_n-\rho_p)/\rho$ with $\rho$, $\rho_n$ and $\rho_p$ being 
the nucleon, neutron and proton density, respectively.
The effective mass defined in Eq. \ref{mstar} is the so-called k-mass which is 
identical to the total effective mass when the single nucleon potential 
is energy-independent\cite{mahaux,neg81}.  
For the single nucleon potential $U_{\tau}$, following Rizzo et al.\cite{riz} 
we use the phenomenological formalism of Bombaci\cite{bom}
\begin{eqnarray}
&&U_\tau(k,u,\delta)=
%\frac{\hbar^2}{2m}k^2+
Au+Bu^\sigma\nonumber\\
&&-\frac{2}{3}(\sigma-1)\frac{B}{\sigma+1}
\umd{3}u^{\sigma}\delta^2
\nonumber\\
&&\pm \qd{-\frac{2}{3}A \umd{0}u -
\frac{4}{3}\frac{B}{\sigma+1}\umd{3}u^{\sigma}\,}\delta\nonumber\\
&& +\frac{4}{5\rz}\qd{\frac{1}{2} (3C-4z_1) \inew{\tau}
+ (C+2z_1)\inew{{\tau^{\prime}}}}\nonumber\\
&&+ \td{C \pm \frac{C-8z_1}{5}\delta}u\cdot g(k),
%\;\;\;\;\;\;\;\;
\label{ibob}
\end{eqnarray}
where $u\equiv \rho/\rho_0$ is the reduced density and $\pm$ 
is for neutrons/protons. In the above $\mathcal{I}_\tau=\itau$, 
$g(k)$ is a momentum regulator $g(k)\equiv 1/\qd{ 1+\td{\frac{k}{\Lambda}}^2 }$, 
and $f_{\tau}(k)$ is the phase space 
distribution function. The parameter $\Lambda=1.5K_F^0$ where $K_F^0$ is the nucleon 
Fermi wave number in symmetric nuclear matter at normal density $\rho_0$. With A=-144 MeV, B=203.3 MeV,
C=-75 MeV and $\sigma=7/6$ the Bombaci formalism reproduces all ground state 
properties including an incompressibility of $K_0$=210 MeV of 
symmetric nuclear matte\cite{bom,riz}. It should be noticed that 
the Bombaci expression is an extension of the well known Gale-Bertsch-Das Gupta 
formalism\cite{gbd} from symmetric to asymmetric nuclear matter. The various terms are motivated
by the HF analysis using the Gogny effective interaction\cite{gale90,das03}. 
This potential depends explicitly on the momentum but not the total energy of nucleons. Thus
the k-mass obtained is the same as the total effective mass. One should also mention 
that only the last term is momentum dependent. Thus the $\Lambda$ parameter sets the scale of the
momentum dependence of the nucleon potential $U_{\tau}$ and also the scale of the 
effective mass. The value of $\Lambda$ parameter was determined by the ground state 
properties of symmetric nuclear matter. It can also appropriately reproduce the momentum dependence of
the empirical isoscalar nuclear optical potential as we shall show in the following. 
The expressions in Eq. \ref{ibob} leads  
to an effective mass\cite{bom,riz}
\begin{equation}
\frac{m^*_\tau}{m_{\tau}}=\left\{1+\frac{-\frac{2m_{\tau}}{\hbar^2} \frac{1}{\Lambda^2}
\td{
 {C \pm \frac{C-8z_1}{5}\delta}}u}{
\left[ 1+ \left( \frac{k_{F0}}{\Lambda}
\right)^{^2}   (1 \pm \delta)^{^{2/3}}
u^{^{2/3}}\right]^2}  \right\}^{-1}.
\end{equation}
It is noticed that the $(1 \pm \delta)^{2/3}u^{2/3}$ term comes from the nuclearn Fermi 
wave number $k_{\tau}^F$ squared, and $\pm$ is for $\tau=n/p$. 
The three parameters $x_0, x_3$ and $z_1$ can be adjusted to mimic different 
predictions on the density-dependent symmetry 
energy and the neutron-proton effective mass splitting. Two sets of parameters can be 
chosen to give two opposite nucleon effective mass splittings, 
but almost the same symmetry energy $E_{sym}(\rho)$\cite{riz}. The parameter set $z_1=-36.75$ MeV, 
$x_0=-1.477$ and $x_3=-1.01$ (case 1) leads to $m_n^*>m_p^*$ while the one with 
$z_1=50$ MeV, $x_0=1.589$ and $x_3=-0.195$ (case 2) leads to $m_n^*<m_p^*$ at all non-zero 
densities and isospin asymmetries. Shown in Fig. 1 are the nucleon effective masses as functions of density 
and isospin asymmetry in both cases. It is seen that the neutron-proton effective mass splittings, 
opposite in signs, increase in magnitudes in both cases 
with the density and isospin asymmetry. Thus large magnitides of effective mass splittings can be obtained 
in dense neutron-rich matter. 
\begin{figure}[ht]
%\centering\epsfig{file=fig1.eps,width=13cm,height=12cm,angle=-90} 
\includegraphics[scale=0.4,angle=-90]{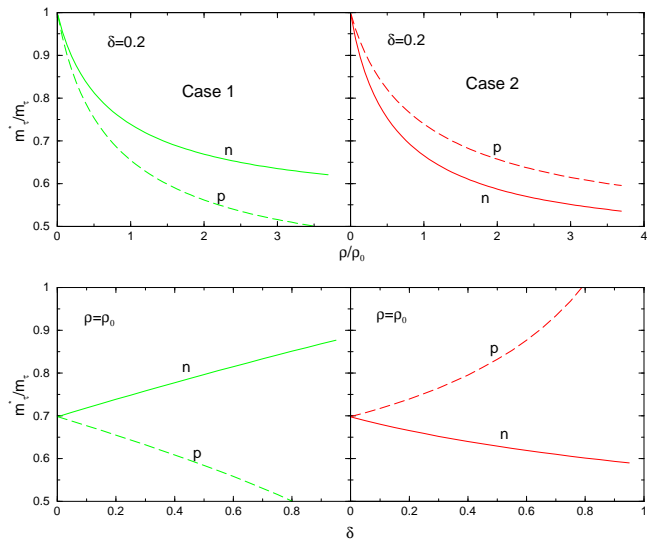}
%\vspace{1.cm}
\caption{{\protect\small (Color on line) Nucleon effective masses as a function 
of density (upper window) and
isospin asymmetry (lower window) with the two parameter sets (see text)}.}
\label{emass}
\end{figure}
These two sets of parameters were shown to lead to rather different results for several observables in 
heavy-ion reactions induced by neutron-rich nuclei\cite{riz}.

\begin{figure}[ht]
%\centering \epsfig{file=fig2.eps,width=13cm,height=12cm,angle=-90} 
\includegraphics[scale=0.55,angle=-90]{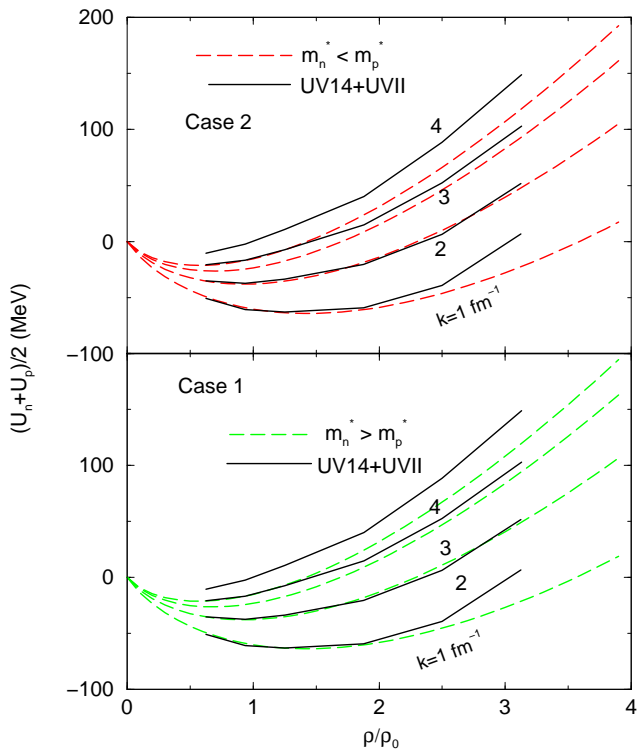}
\vspace{0.6cm}
\caption{{\protect\small (Color on line) Strengths of the isoscalar potential as 
a function of density at the four wave numbers for case 1 (lower window) 
and case 2 (upper window) in comparison with the variational 
many-body calculations }}
\end{figure}
Our analysis is based on the consideration that both the isoscalar and isovector parts of 
the single nucleon potential must have asymptotic values at $\rho_0$ in agreement with 
the real parts of the corresponding nucleon optical potentials constrained by nucleon-nucleus 
scattering experiments. We first examine the isoscalar potentials with the two sets of parameters. 
The quantity $(U_n+U_p)/2$ is a 
good approximation of the isosclar potential since the $\delta^2$ in eq. \ref{ibob} 
is negligibly small. It is now customary and 
a more stringent test of the isoscalar potentials to compare them with the 
variational many-body (VMB) predictions by 
Wiringa\cite{wiringa88,gale90,pawel91,dan02}. In the VMB theory the single 
nucleon potential was obtained by using a realistic Hamiltonian that fits nucleon-nucleon 
scattering data, few-body nuclear binding energies and nuclear matter saturation 
properties. It also reproduces the experimental nucleon optical potential available mainly at low 
energies\cite{fai}. Shown in Fig.\ 2 is a comparison of isosclar potentials using eq. \ref{ibob} 
with the VMB predictions using the $UV14$ two-body potential and the $UVII$ three-body 
potential\cite{wiringa88}. First of all, the isoscalar potentials are almost 
independent of the neutron-proton effective mass splittings as one expects. In both cases the 
isoscalar potentials using eq. \ref{ibob} are in good agreement with the VMB predictions 
up to about $k=2.5fm^{-1}$. At higher momenta where combinations of 
different two-body and three-body forces lead to somewhat different predictions 
especially at high densities\cite{wiringa88}, the Bombaci formalism leads to slightly 
lower values. Nevertheless, the quality of agreement with the VMB 
predictions found here is compatible with those using all other models\cite{gale90,pawel91}.   

We now turn to the isovector part of the nucleon potential in both cases. 
The strength of the isovector nucleon optical potential, i.e., the symmetry or Lane potential\cite{lane},
can be extracted from $U_{Lane}\equiv (U_n-U_p)/2\delta$ at $\rho_0$. 
Systematic analyses of a large number of nucleon-nucleus scattering experiments 
at beam energies below about 100 MeV\cite{data} indicates undoubtedly that 
the Lane potential decreases approximately linearly with increasing 
beam energy $E_{kin}$, i.e.,  
\begin{equation}\label{dat}
U_{Lane}=a-bE_{kin}
\end{equation}
where $a\simeq 22-34$ MeV and $b\simeq 0.1-0.2$\cite{theory1,theory2}.  
\begin{figure}[ht]
%\centering \epsfig{file=fig3.eps,width=13cm,height=12cm,angle=-90} 
\includegraphics[scale=0.45,angle=-90]{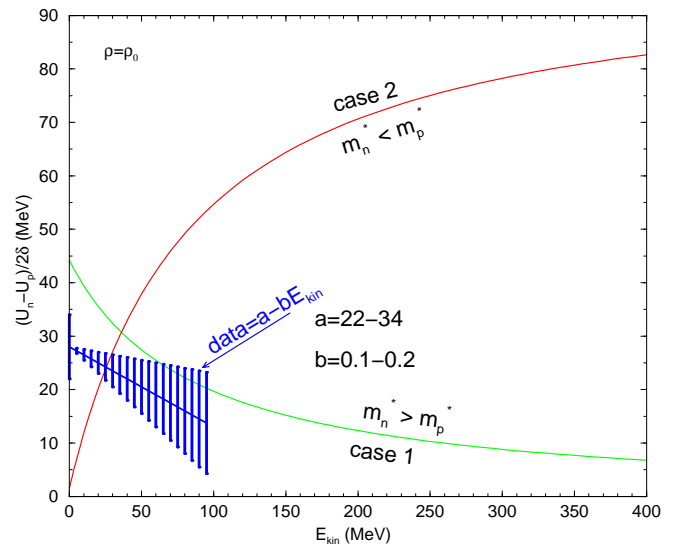}
%\vspace{1.cm}
\caption{{\protect\small (Color on line) Strength of the isovector potential at 
normal density $\rho_0$ as a function of nucleon kinetic energy.}}
\end{figure}
Shown in Fig.\ 3 are the theoretical symmetry potentials in the two cases in comparison 
with the Lane potential constrained by the experimental data. 
The vertical bars are used to indicate the 
uncertainties of the coefficients $a$ and $b$ in eq. \ref{dat}. 
It is seen that with the effective mass splitting $m_n^*>m_p^*$ (case 1) the strength of the 
symmetry potential decreases with increasing energy. This trend is in agreement with the 
experimental indication. Moreover, the slope of the calculated symmetry potential with respect to energy 
is about right although the magnitude obtained is slightly higher. Since it is not the purpose of 
this work to find the best parameter set to reproduce the empirical Lane potential, no attempt is made to 
readjust any of the parameters first proposed in refs.\cite{bom,riz}. 
In case 2, however, the most striking feature is that
the symmetry potential increases with the increasing beam energy. 
This is in stark contrast with the experimental indications. The characteristically wrong 
energy dependence of the symmetry potential in this case thus excludes definitely the neutron-proton 
effective mass splitting of $m_n^*<m_p^*$ in neutron-rich matter.   

Our finding here is consistent with that of the earlier work by Sj\"oberg 
in the framework of the Landau-Fermi liquid theory\cite{sjo76}, BHF predictions\cite{bhf} as well 
as Hartree-Fock calculations using the Gogny effective interaction\cite{lidas}. 
Within the Landau-Fermi liquid theory, very interestingly, the nucleon effective mass 
splitting has the analytical and physically transparent relation\cite{sjo76,ref}
\begin{eqnarray}\label{sjo}
(m_n^* - m_p^*)/m &=& \frac{m_n^* k_n}{3 \pi^2} \left[ f_1^{nn} + (k_p/k_n)^2 f_1^{np} \right]\\\nonumber
                &-&\frac{ m_p^* k_p}{3 \pi^2} \left[ f_1^{pp} + (k_n/k_p)^2 f_1^{np} \right],
\end{eqnarray} 
where $f_1^{nn},f_1^{pp}$ and $f_1^{np}$ are the neutron-neutron, proton-proton and neutron-proton 
quasiparticle interactions projected on the $l=1$ Legendre polynomial, as for the effective 
mass in a one-component Fermi liquid. Microscopic NN interactions predict all $f_1$'s are negative 
in symmetric matter at nuclear matter density and in asymmetric matter at tree-level. 
It can be seen then that the proton effective mass is smaller in neutron-rich matter 
due to the coupling of the protons to the denser neutron background, i.e., the term 
$(k_n/k_p)^2 f_1^{np}$ is dominant in the Eq.\ref{sjo} above, leading to $m_n^*>m_p^*$ as shown
numerically in Fig.\ 3 of ref.\cite{sjo76}.

In summary, using the Bombaci phenomenological formalism for single nucleon potentials 
in isospin asymmetric nuclear matter we analyzed the isoscalar and isovector 
nucleon optical potentials in comparison with the empirical ones constrained by the 
experimental data. We found that a neutron-proton effective mass splitting of $m_n^*< m_p^*$ 
leads to a characteristically wrong energy dependence of the Lane potential. 
Thus, the possibility of $m_n^*< m_p^*$ in neutron-rich matter is ruled out indirectly by 
the experimental data.

This research is supported in part by the National Science Foundation 
under grant No. PHY-0088934 and PHY-0243571.

\end{document}